\documentclass[10pt]{article}
\usepackage[utf8]{inputenc}
\usepackage[T1]{fontenc}
\usepackage[a4paper,margin=1.9cm]{geometry}
\usepackage{authblk}
\usepackage{float}
\usepackage{graphicx}
\usepackage{amsmath,amssymb}
\usepackage[font=small,labelfont=bf,labelsep=period]{caption}
\usepackage[numbers,sort&compress,super]{natbib}
\usepackage[colorlinks=true,citecolor=blue,linkcolor=blue,urlcolor=blue]{hyperref}

\makeatletter
\renewcommand{\maketitle}{%
  \begin{center}
    {\LARGE\bfseries \@title \par}
    \vspace{1.0em}
    {\normalsize \@author \par}
  \end{center}
  \vspace{1.0em}
}
\makeatother

\renewenvironment{abstract}{%
  \begin{center}
  \begin{minipage}{0.86\textwidth}
  \small
}{%
  \end{minipage}
  \end{center}
  \vspace{1.0em}
}

\title{Generation of dense relativistic electron beams via vortex laser-driven self-generated magnetic pinching}

\author[1]{Mingxuan Wei\thanks{These authors contributed equally to this work.}}
\author[2,3]{Fengyu Sun$^{*}$}
\author[4]{Zhongpeng Li$^{*}$}
\author[1]{Xichen Hu}
\author[2]{Huiting Ma}
\author[1]{Guangwei Lu}
\author[1]{Zhuofan Zhang}
\author[4]{Lijie Cui}
\author[2,3]{Qijin Zhang}
\author[2,3]{Mengjiao Wang}
\author[1]{Weijun Zhou}
\author[4]{Qian Zhao}
\author[4]{Wenqing Wei}
\author[2]{Yi Xu}
\author[2]{Zongxin Zhang}
\author[2]{Jiayi Qian}
\author[2]{Jiacheng Zhu}
\author[2,5]{Xiaoyan Liang}
\author[1]{Min Chen}
\author[2]{Wenpeng Wang$^\dagger$}
\author[4,6]{Jian-Xing Li$^\dagger$}
\author[1]{Wenchao Yan$^\dagger$}
\author[2,5]{Yuxin Leng}
\author[1,7]{Jie Zhang}

\affil[1]{State Key Laboratory of Dark Matter Physics, Key Laboratory for Laser Plasmas (MoE), School of Physics and Astronomy, Shanghai Jiao Tong University, Shanghai 200240, China}
\affil[2]{State Key Laboratory of High Field Laser Physics and CAS Center for Excellence in Ultra-intense Laser Science, Shanghai Institute of Optics and Fine Mechanics (SIOM), Chinese Academy of Sciences (CAS), Shanghai 201800, China}
\affil[3]{University of Chinese Academy of Sciences, Beijing 100049, China}
\affil[4]{Ministry of Education Key Laboratory for Nonequilibrium Synthesis and Modulation of Condensed Matter, State Key Laboratory of Electrical Insulation and Power Equipment, Shaanxi Province Key Laboratory of Quantum Information and Quantum Optoelectronic Devices, School of Physics, Xi'an Jiaotong University, Xi'an 710049, China}
\affil[5]{School of Physical Science and Technology, ShanghaiTech University, Shanghai 201210, China}

\affil[6]{Department of Nuclear Physics, China Institute of Atomic Energy, P.O. Box 275(7), Beijing 102413, China}
\affil[7]{Tsung-Dao Lee Institute, Shanghai Jiao Tong University, Shanghai 201210, China}

\affil[ ]{\textsuperscript{\ensuremath{\dagger}}Correspondence to: \texttt{wangwenpeng@siom.ac.cn}; \texttt{jianxing@xjtu.edu.cn}; \texttt{wenchaoyan@sjtu.edu.cn}}
\affil[ ]{\textsuperscript{\ensuremath{*}} These authors contributed equally to this work.}

\begin{document}

\flushbottom
\maketitle
\begin{abstract}
In multi-petawatt laser plasma accelerators, achieving high-density relativistic electron beams is typically accompanied by large transverse divergence, limiting the attainable effective electron density needed to access high-flux interaction regimes relevant to laboratory astrophysics.
Here we report experimental demonstration of self-generated magnetic pinching (SMP), a collective mechanism that actively regulates transverse beam dynamics using a Laguerre–Gaussian laser at strong relativistic intensity ($\sim 8 \times 10^{19}~\mathrm{W/cm^2}$) interacting with an underdense plasma. 
The electron beam evolves from a two-lobe high-charge injection structure into a compressed, high-density profile, yielding a threefold reduction in divergence and nearly an order-of-magnitude enhancement in effective beam density compared with a Gaussian driver.
Particle-in-cell simulations agree well with the experimental observations and reveal that a self-generated azimuthal magnetic field governs the electron dynamics within the SMP regime, which is defined by the forming condition
\(S \approx 0.717\,\ell a_0\,\left[n_e(10^{18}~\mathrm{cm^{-3}})\right]^{-3/4}\approx1\),
where \(\ell\), \(a_0\), and \(n_e\) are topological charge, laser amplitude, and plasma density, respectively. A transient kick from a dense inner sheath electron population drives collective magnetic pinching, transforming an initially separated electron distribution into a compressed and well-collimated beam.
For higher-power laser systems, the forming condition can be extended to higher plasma densities and larger orbital angular momentum modes, potentially enabling electron beams with charges exceeding several nC and effective densities above $10^{19}~\mathrm{cm^{-3}}$, surpassing those achievable with current multi-petawatt systems. 
This mechanism provides a route to overcoming transverse expansion, enabling significantly enhanced rare interaction probabilities, including high-flux neutron generation and correlated two-neutron capture processes relevant to heavy-element nucleosynthesis.

\end{abstract}

%
%
\thispagestyle{empty}

\section*{Main}
Laser plasma acceleration (LPA) can generate relativistic electron beams over millimetre-scale distances by sustaining accelerating gradients at the teravolt-per-metre level. This capability has opened opportunities in high-energy-density physics, compact radiation sources and laboratory astrophysics\cite{tajimaLaserElectronAccelerator1979, esareyPhysicsLaserdrivenPlasmabased2009,schlenvoigtCompactSynchrotronRadiation2008,wuHighthroughputInjectionAcceleration2021}. 
Driven by intense laser pulses with Gaussian spatial profiles, plasma wakes can form linear or nonlinear bubble structures that confine and accelerate electrons near the axis \cite{luNonlinearTheoryMultidimensional2006}. 
Over the past two decades, significant progress has been made in improving beam energy~\cite{poderMultiGeVElectronAcceleration2024,picksleyMatchedGuidingControlled2024}, stability \cite{winklerActiveEnergyCompression2025} and energy spread~\cite{wangHighBrightnessHighEnergyElectron2016,keNearGeVElectronBeams2021a}, driven by advances in laser technology and plasma control~\cite{bjorklundsvenssonLowdivergenceFemtosecondXray2021}. 
However, a fundamental limitation persists: achieving high beam density is typically accompanied by increased transverse divergence, whereas well-collimated beams generally contain only modest charge.
This reflects an intrinsic trade-off between beam charge and transverse confinement, posing a central challenge for simultaneously achieving high charge and low divergence in LPA, which are critical for laboratory astrophysics and high-flux nuclear interaction studies.

This trade-off is closely linked to the underlying acceleration mechanisms, which operate in distinct regimes defined by plasma density and interaction dynamics. In laser-driven acceleration, laser wakefield acceleration (LWFA), typically realized in low-density plasmas~\cite{faureLaserPlasmaAccelerator2004,geddesHighqualityElectronBeams2004,manglesMonoenergeticBeamsRelativistic2004}, produces high-energy, well-collimated electron beams, but is often limited by beam loading effects~\cite{couperusDemonstrationBeamLoaded2017a}. In contrast, direct laser acceleration (DLA)~\cite{pukhovParticleAccelerationRelativistic1999a}, which becomes more prominent at higher plasma densities~\cite{gahnMultiMeVElectronBeam1999a,shawRoleDirectLaser2017a,babjakDirectLaserAcceleration2024a}, enables substantially higher charge extraction through direct coupling between the laser field and electrons, but typically at the cost of increased transverse divergence~\cite{huHundredsNanocoulombElectron2025}.
Although several approaches, including plasma lenses~\cite{pompiliFocusingHighbrightnessElectron2018,changReductionElectronbeamDivergence2023}, density tailoring~\cite{fangSharpPlasmaPinnacle2018}, and driver shaping~\cite{cormier-michelControlFocusingFields2011}, have been developed to mitigate beam divergence, these methods generally rely on relatively stable or quasi-static wake structures. This requirement becomes increasingly difficult to satisfy in ultra-intense multi-petawatt laser systems, where practical engineering constraints typically impose short-focal configurations because the long-focal-length geometries required for matching conditions are difficult to realize~\cite{zengSimulationObservationHigh2024,luNonlinearTheoryMultidimensional2006}. Under such strongly nonlinear conditions, the plasma wake evolves rapidly, leading to enhanced transverse momentum growth and beam spreading during acceleration~\cite{pukhovLaserWakeField2002}. Consequently, achieving simultaneous high charge and low divergence remains a major challenge, highlighting the need for active control of transverse beam dynamics directly during the acceleration process.

Structured light may provide a potential solution to overcoming this challenge, owing to its intrinsic phase gradients and orbital angular momentum (OAM), which have been widely exploited for manipulating microscopic particles and complex matter in the non-relativistic regime~\cite{patersonControlledRotationOptically2001a,rosales-guzmanReviewComplexVector2018a,shenOpticalVortices302019a}.
In laser plasma interactions, Laguerre–Gaussian (LG) laser pulses have been shown to modify plasma structures and enhance proton acceleration and collimation\cite{wangHollowPlasmaAcceleration2020,wangProtonAccelerationDriven2024}.
For electron acceleration, LG pulses can drive doughnut-shaped wakefields that support hollow or ring-shaped beams \cite{zhangAccelerationEvolutionHollow2016,zhangAccelerationOnaxisRingshaped2016,vieiraNonlinearLaserDriven2014}. Recent work has experimentally demonstrated that relativistic LG laser pulses can generate directional electron beams in the DLA regime \cite{wangExperimentalDemonstrationDirectional2026}, where beam collimation originates from direct coupling between relativistic electrons and the reflected structured laser field rather than from collective plasma dynamics. These studies highlight the additional degrees of freedom introduced by transverse optical modes for manipulating relativistic electrons. However, whether optical mode engineering can mitigate the long-standing trade-off in nonlinear wakefield acceleration by regulating the collective evolution of plasma dynamics, instead of relying on direct laser--electron interactions, remains unknown.

\begin{figure}[h!]
    \centering
    \includegraphics[width=0.8\linewidth]{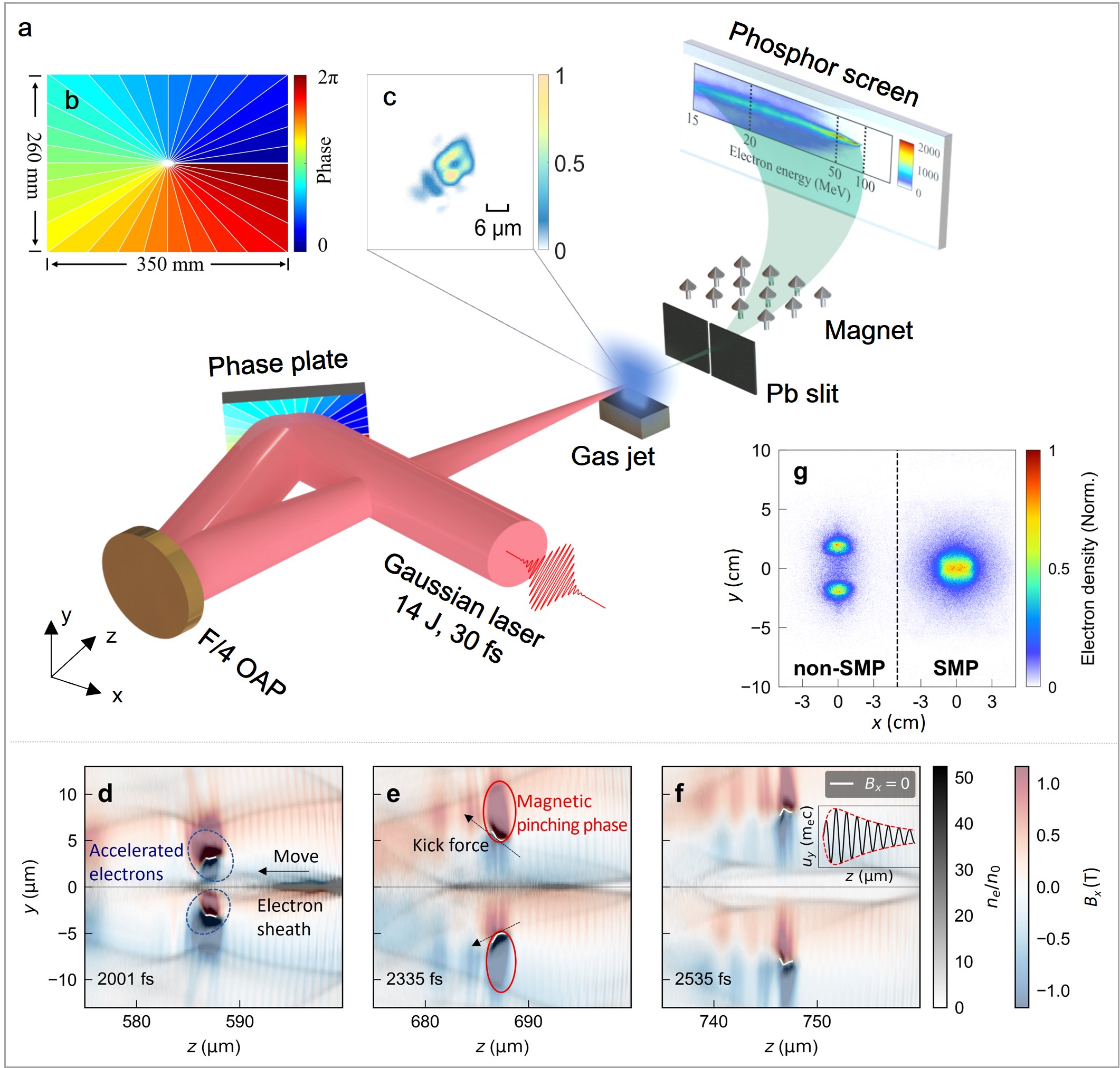}
    \caption{\textbf{a}, Configuration of the experimental setup (not to scale). A linearly polarized Gaussian laser pulse (14 J, 30 fs) was focused by an $F/4$ off-axis parabolic mirror onto a pure nitrogen gas jet, with the beam propagating along the $+z$ direction. A spiral phase plate with topological charge $\ell=1$ was placed before the mirror to convert the Gaussian beam into a LG mode. The accelerated electrons passed through a lead slit and were dispersed by a dipole magnet onto a phosphor screen for charge and energy measurements. Panel \textbf{b} displays the phase distribution of the spiral phase plate. Panel \textbf{c} shows the focal spot of the LG laser. \textbf{d}--\textbf{f}, Simulated evolution of the nonlinear doughnut wake and the associated self-generated magnetic field during electron acceleration. The greyscale maps show the electron density, while the red--blue overlay represents the self-generated transverse magnetic field $B_x$ (normalized). The white curves mark the local $B_x=0$ phase boundary near the accelerated electron bunch.  At the early stage (\textbf{d}), the doughnut  wake contains accelerated electrons in the doughnut bubble and background electrons in the inner sheath. As the wake evolves, the inner sheath moves toward the axis and imparts a transverse kick to the accelerated bunch. This kick drives the bunch into the magnetic pinching phase (red solid-line ellipse in \textbf{e}), where a large fraction of electrons cross the $B_x=0$ boundary, as shown in \textbf{e} and \textbf{f}. Once loaded into this phase, the self-generated magnetic field continuously removes transverse momentum $u_y$ through the $\mathbf{v}\times\mathbf{B}$ interaction, resulting in sustained magnetic pinching, as illustrated by the gradual decrease of the transverse amplitude (red line) in the inset of \textbf{f}. The resulting collective magnetic pinching produces the compressed beam observed in the far-field profiles (\textbf{g}), in excellent agreement with experimental observations. Electron densities are normalized to their respective maxima.}
    \label{fig1}
\end{figure}

Here we demonstrate the experimental verification of self-generated magnetic pinching (SMP)  of relativistic electron
beams in twisted-laser plasma acceleration, a collective mechanism driven by a relativistic LG laser that actively regulates transverse dynamics and induces beam compression. When the laser amplitude and plasma density satisfy the forming condition \(S \approx 0.717\,\ell a_0\,\left[n_e(10^{18}~\mathrm{cm^{-3}})\right]^{-3/4}\approx1\), the interaction enters an intermediate density regime, in which the plasma density exceeds that typically employed in LWFA while remaining well below the near-critical densities characteristic of DLA. Under these conditions, the LG-driven wake supports the formation of a high-charge two-lobe relativistic electron beam, whose structured transverse current distribution generates sufficiently strong quasi-static azimuthal magnetic fields, as illustrated in Fig.~\ref{fig1}d. 
During propagation, the tightly focused LG pulse drives a nonlinear evolution of the doughnut-like wake in which the inner ring of the bubble merges on axis. At the same time, the inner electron sheath undergoes a defocusing evolution, generating a transverse impulse on the accelerated bunch. This kick drives the bunch into the self-generated magnetic pinching phase, during which the self-generated magnetic field reduces the transverse momentum through the $\mathbf{v}\times\mathbf{B}$ force while simultaneously inducing collective magnetic pinching, thereby compressing the electron bunch, as shown in Fig.~\ref{fig1}e and Fig.~\ref{fig1}f.
SMP thus represents a distinct regime in which the laser mode structure actively controls beam dynamics during acceleration, rather than merely shaping the wakefield. 

\section*{Observation of electron beam compression}

The experiment was conducted on the 1~PW laser system at the SULF facility~\cite{zhangLaserBeamlineSULF2020}, with an on-target power of 0.5~PW, as described in detail in the Methods.  
The 800-nm laser pulse was focused by an F/4 off-axis parabola onto a 1.2-mm pure nitrogen gas jet as shown in Fig.~\ref{fig1}a, operating in the ionization injection regime~\cite{claytonSelfguidedLaserWakefield2010a,pakInjectionTrappingTunnelionized2010a,chenTheoryIonizationinducedTrapping2012}. 
Under this short-focal-length condition, a Gaussian driver is focused to a $6.6~\mu$m (FWHM) spot, corresponding to a peak intensity of $\sim3.4\times10^{20}~\mathrm{W/cm^2}$, yielding a normalized laser amplitude
$a_0=0.855\sqrt{I_0/(10^{18}~\mathrm{W/cm^2})}\,(\lambda_0/\mu\mathrm{m})\approx12$,
where $I_0$ is the peak laser intensity. Such a tightly focused, ultra-intense driver excites a strongly nonlinear wakefield and induces rapid evolution of the plasma bubble.
As a result, electron injection is enhanced and large beam charge can be achieved, but the accelerated electrons experience strong transverse forces and develop substantial divergence. 

\begin{figure}[htbp]
    \centering
    \includegraphics[width=0.8\linewidth]{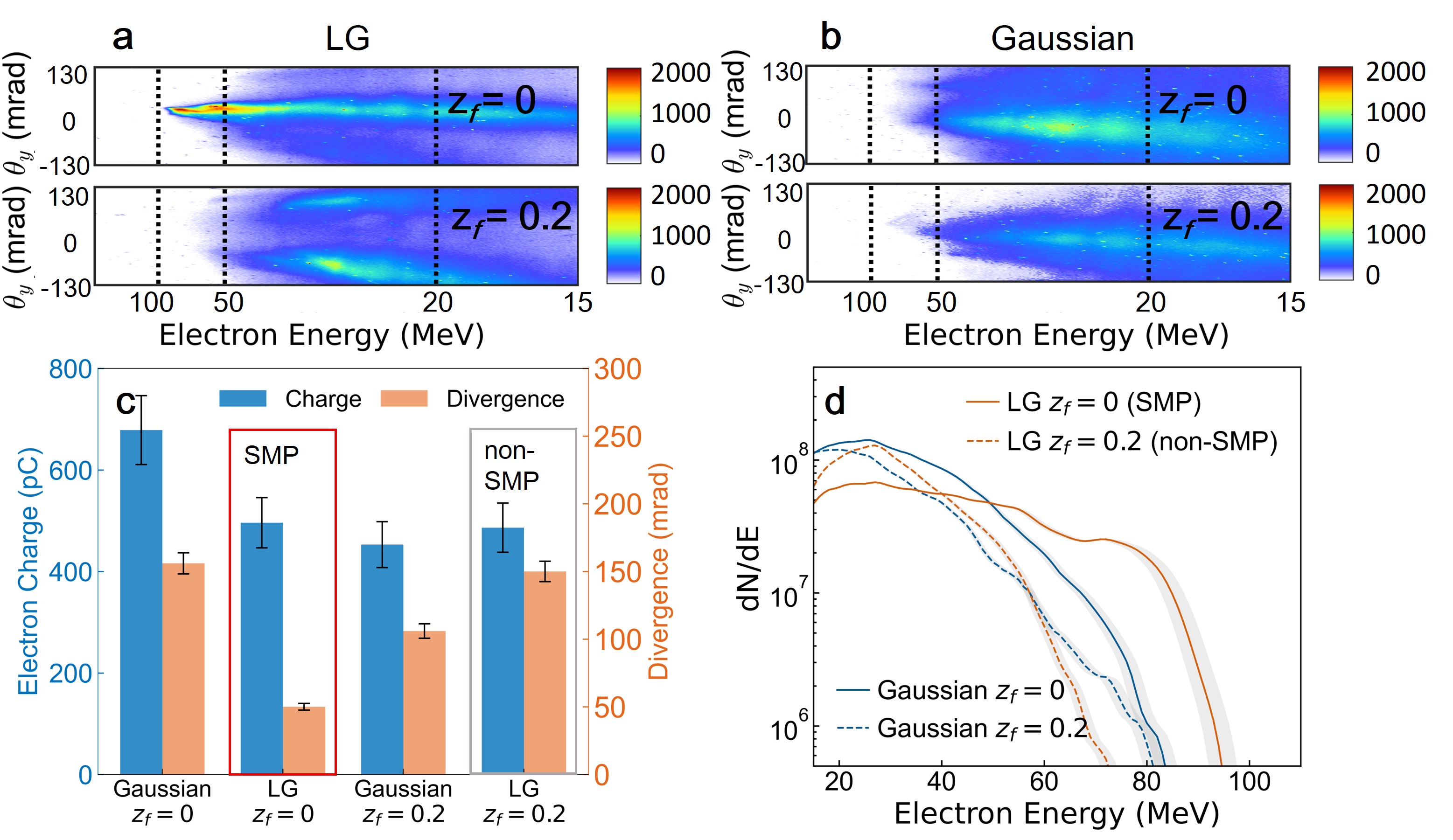}
    \caption{Typical energy spectra and characteristics of accelerated electron beams driven by Gaussian and LG lasers at different focal positions $z_f$. \textbf{a,b}, Measured electron spectra for \textbf{(a)} LG and \textbf{(b)} Gaussian lasers at $z_f = 0$ and 0.2~mm. \textbf{c}, Comparison of electron charge (blue) and divergence $\theta_y $(orange) for both laser configurations, where $\theta_y = \arctan (u_y/u_z)$ evaluated at the FWHM. The red box highlights the LG case with $z_f=0$, where the formation of an SMP results in a significant reduction in divergence. \textbf{d}, Corresponding energy spectra integrated at $z_f$ = 0 and $z_f$ = 0.2~mm; the gray shaded area represents the error bars.}
    \label{fig2}
\end{figure}

Figure \ref{fig2} summarizes the electron beams accelerated by the LG and Gaussian laser, respectively. For the LG case, the angular distribution depends strongly on the focal position, as shown in Fig.~\ref{fig2}a. At $z_f = 0.2~\mathrm{mm}$, the beam exhibits a pronounced two-lobe structure oriented perpendicular to the laser polarization. As the focus is shifted to $z_f = 0$,  corresponding to the SMP condition for the LG driver, this structured distribution progressively contracts and eventually transforms into a narrow, well-collimated beam. In contrast, the Gaussian case shows much weaker dependence on $z_f$, and the electrons remain broadly distributed without a distinct collimation transition as shown in Fig.~\ref{fig2}b. The quantitative comparison in Fig.~\ref{fig2}c shows that, for the Gaussian laser, the electron beam with energies above 
$15~\mathrm{MeV}$ carries about $678~ \pm 45~\mathrm{pC} $ at $z_f = 0$ with a divergence of $156~ \pm 8~\mathrm{mrad}$. 
When the LG laser is applied at the same focal position, the beam divergence
shrinks from $156 \pm 8$ mrad to $50 \pm 3$ mrad, representing a threefold
improvement in collimation, while the charge decreases from $678 \pm 45$ pC
to $497 \pm 27$ pC (a reduction of approximately 27\%). These two changes
act in opposite directions: the charge reduction penalizes total-yield-driven
applications, whereas the collimation improvement benefits flux-density-driven
applications. To quantify the net effect on the latter, we define an
effective electron density $n_{\mathrm{eff}}$ that accounts for both charge
and transverse divergence at a fixed interaction plane (see below). For the
LG-driven beam at the SMP condition, we obtain $n_{\mathrm{eff}} \approx
4.5 \times 10^{18}$ cm$^{-3}$, compared with $n_{\mathrm{eff}} \approx
6.4 \times 10^{17}$ cm$^{-3}$ for the Gaussian-driven beam---an approximately
sevenfold enhancement. The gain in collimation thus more than compensates
for the modest charge reduction. We note that part of the charge difference
between the Gaussian and LG cases may be attributable to the reduced
effective $a_0$ of the LG pulse at fixed incident energy; a Gaussian driver
with comparable $a_0 \approx 6$ would operate in a substantially different
wakefield regime and is not directly comparable under the present
experimental conditions.

\begin{figure}[h]
    \centering    \includegraphics[width=0.8\linewidth]{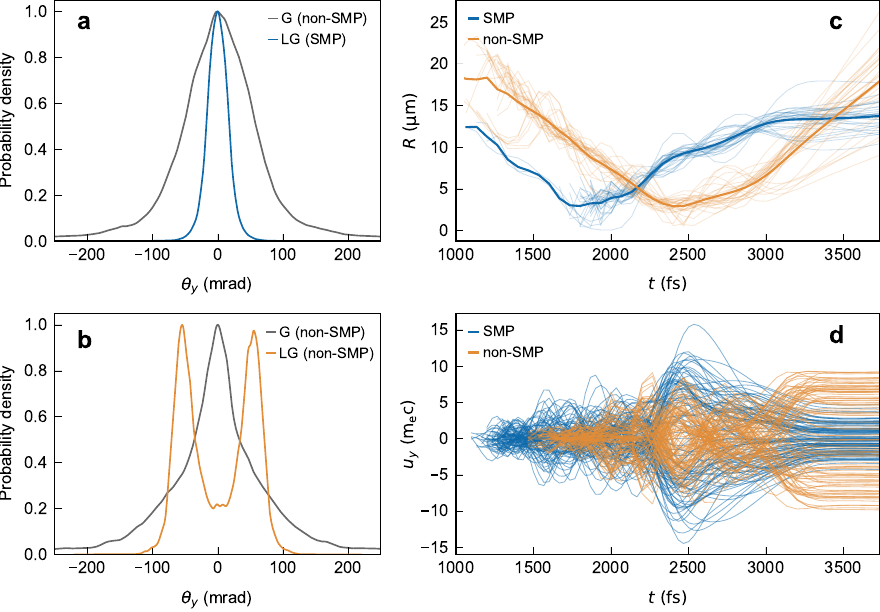}
    \caption{Comparison of the divergence and transverse dynamics of accelerated electrons driven by Gaussian and LG laser. Simulated electron divergence at two focal positions, \textbf{(a)} $z_f = 0~\mathrm{mm}$ and \textbf{(b)} $z_f = 0.2~\mathrm{mm}$ for different driver laser configurations.
    \textbf{c,} Temporal evolution of the transverse off-axis radius $R(t)$ of the accelerated electrons driven by LG pulses for the case with and without SMP. The thin curves represent trajectories of 50 randomly selected electrons, while the thick lines denote their average values. \textbf{d,} Temporal evolution of the transverse momentum $u_y(t)$ for the electrons selected in \textbf{(c)}, illustrating their oscillatory motion during acceleration. }
    \label{fig3}
\end{figure}

For a plasma electron density of $n_e = 7.0\times10^{18}~\mathrm{cm^{-3}}$, 
the characteristic accelerating gradient of the wakefield can be estimated from $E_{\mathrm{0}}(\mathrm{GV/m}) \approx 96~\sqrt{{n_e}({10^{18}~\mathrm{cm^{-3}}})}$,
which yields $E_{\mathrm{0}} \approx 250~\mathrm{MeV/mm}$ under our plasma conditions\cite{esareyPhysicsLaserdrivenPlasmabased2009,dipiazzaExtremelyHighintensityLaser2012}. In the simulations, for the Gaussian laser at $z_f = 0$, high energy electrons experience acceleration primarily over a distance from $z \approx 220~\mu\mathrm{m}$ to $580~\mu\mathrm{m}$, 
corresponding to an effective acceleration length of about $360~\mu\mathrm{m}$. 
The resulting maximum energy reaches approximately $90~\mathrm{MeV}$, consistent with the expected LWFA energy gain. 
When the focal position is adjusted to the SMP condition for the LG driver, the maximum electron energy increases to about $100~\mathrm{MeV}$, as shown in Fig.~\ref{fig2}d. This enhancement indicates that the SMP configuration alters the acceleration dynamics, resulting in additional energy gain relative to the case without SMP.
Together, these results demonstrate that the SMP condition simultaneously improves beam collimation and enhances overall acceleration performance.

\section*{Collective electromagnetic dynamics driven by SMP}
\textit{PIC simulations reveal beam evolution.}
As shown in Fig.~\ref{fig3}a, the quasi-3D PIC simulations~\cite{leheSpectralQuasicylindricalDispersionfree2016} reproduce the main features of the experimentally observed divergence evolution. Compared with the Gaussian driver, the LG laser under the SMP condition generates an electron beam with a full width at half maximum (FWHM) divergence reduced by a factor of about 3, reaching 45~mrad, while the simulated beam charge is $535~\mathrm{pC}$, in good agreement with the experimental measurement.
The simulations also capture the characteristic two-lobe electron structure driven by the LG-induced doughnut wakefield \cite{zhangAccelerationEvolutionHollow2016,zhangAccelerationOnaxisRingshaped2016}. The angular modulation of the transverse field in a linearly polarized LG pulse breaks cylindrical symmetry and imprints polarization-dependent transverse momentum on newly ionized electrons. Because trapping in the nonlinear wake requires small radial momentum, electrons born perpendicular to the polarization axis are preferentially confined. As shown in Supplementary Fig. S1, this selective trapping results in a pronounced two-lobe injection structure. The resulting electron angular distribution exhibits a clearly symmetric double-peaked structure, as illustrated in Fig.~\ref{fig3}b. Corresponding to this is the temporal evolution of the electron mean off-axis radius $R$ shows an initial contraction followed by expansion, reflecting the alternating focusing and defocusing phases in the LG driven wakefield. In contrast, under the SMP condition, the radius stabilizes after $t \approx 3000~\mathrm{fs}$, indicating the formation of a quasi-collimated beam rather than a two-lobe structure that continues to expand with an approximately constant slope, as shown in Fig.~\ref{fig3}c.
This transition is further supported by the evolution of the transverse momentum $u_y$ in Fig.~\ref{fig3}d, where the oscillation amplitude of $u_y$ decreases markedly around $t \approx 3000~\mathrm{fs}$, demonstrating that the electrons become confined within the beam core, as detailed in Fig.~\ref{fig1}g. A direct comparison between simulations and far-field measurements, showing excellent agreement, is provided in the Supplementary Fig.~S3.

\begin{figure}[h]
    \centering
    \includegraphics[width=0.8\linewidth]{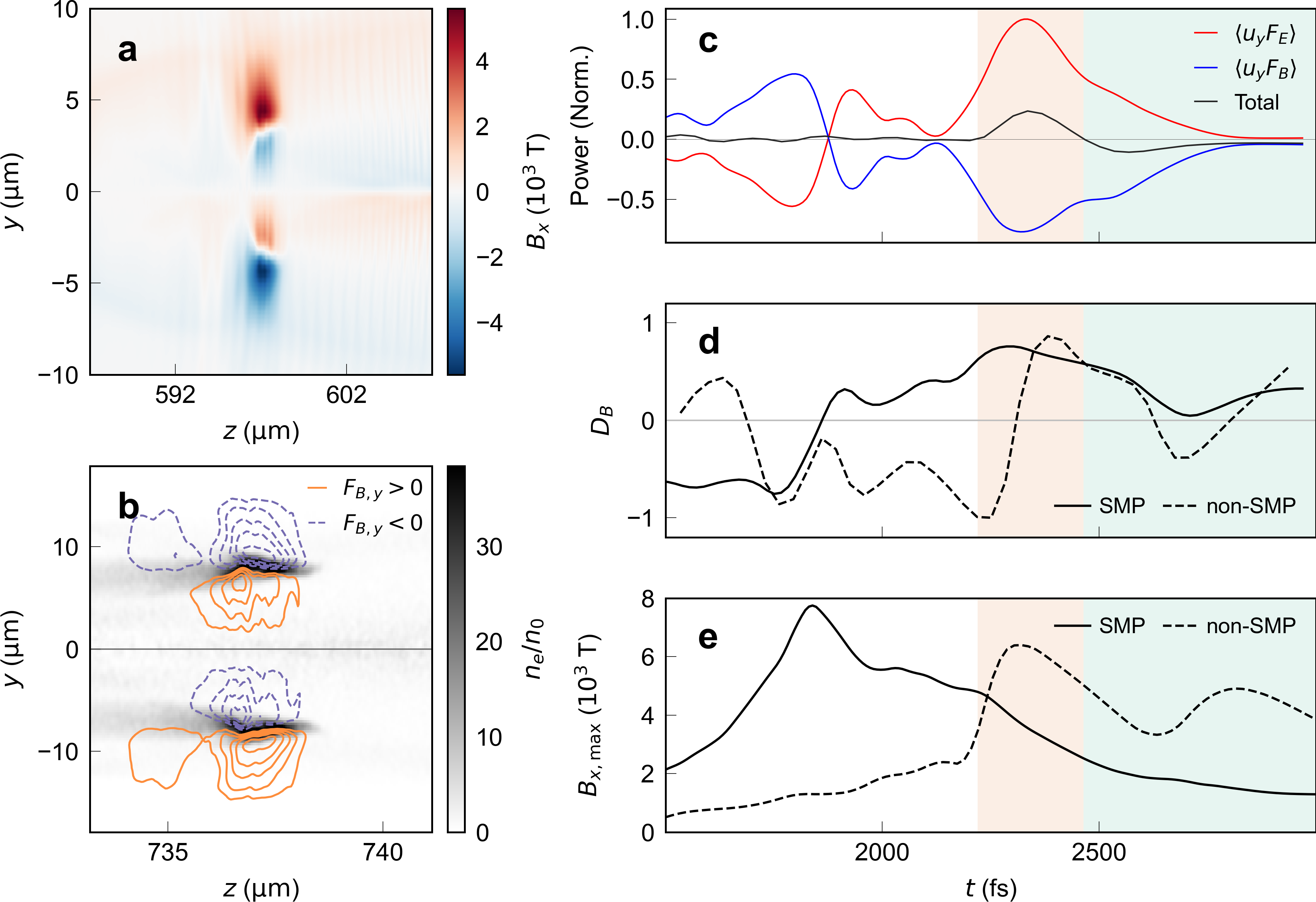}
    \caption{\textbf{a,} Spatial distribution of the quasi-static azimuthal magnetic field $B_x$ generated by the plasma current driven by the LG laser. \textbf{b,} Electron density $n_e/n_0$ overlaid with contours of the transverse magnetic force per charge, $F_{B,y}/e=-v_zB_x$. The upper and lower electron lobes predominantly overlap with regions where the magnetic force is directed toward the axis, indicating the onset of magnetic pinching. \textbf{c,} Temporal evolution of the electric (red), magnetic (blue), and total (black) contributions to the transverse power. \textbf{d,} Particle-weighted magnetic pinching index $D_B(t)$ for the case with and non-SMP under the LG case. Positive values correspond to net magnetic pinching of the transverse motion. \textbf{e,} Temporal evolution of $B_{x,\max}$ for the cases with and non-SMP. The SMP configuration sustains a stronger self-generated magnetic field over a longer duration, enabling efficient magnetic pinching and subsequent beam compression.}
    \label{fig4}
\end{figure}

\textit{Analysis of SMP mechanism.}
To elucidate the observed beam collimation, we analysed the coupling between transverse electron motion and the self-generated electromagnetic fields. Figure~\ref{fig4}a  shows the spatial distributions of the azimuthal magnetic field $B_x$. 
A pair of anti-symmetric $B_x$ lobes forms in the wakefield, corresponding to opposite magnetic polarities above and below the axis\cite{shiMagneticFieldGeneration2018b,shiGenerationUltrarelativisticMonoenergetic2021}. 
The formation of this field originates from the two-lobe electron current distribution $\mathbf{J}$ induced by the LG laser\cite{shiElectronPulseTrain2022,shiMagneticFieldGeneration2018b}.  
In addition, the helical phase structure of the LG pulse induces finite OAM components in a fraction of the plasma electrons, leading to spiral trajectories observed in the simulations. These combined transverse and azimuthal current components sustain a quasi-static magnetic field $B_x$ determined by  
$\nabla\times\mathbf{B}=\mu_0\mathbf{J}+{1}/{c^2}{\partial\mathbf{E}}/{\partial t}$,
where the ${\partial\mathbf{E}}/{\partial t}$ term is negligible and the plasma current primarily sustains the $B_x$ structure.

The transverse dynamics can be quantified by decomposing the evolution of the transverse momentum into the work performed by the electric and magnetic fields based on the relativistic equation of motion. Figure~\ref{fig4}c shows the electric, magnetic and total contributions to the transverse power. The total curve follows the evolution of $u_y$ in Fig.~\ref{fig3}d, confirming that the power decomposition captures the main transverse dynamics. 
For $t<2220~\mathrm{fs}$, the total power oscillates around zero, indicating that the transverse electric-field driving and magnetic-field pinching approximately balance each other. Correspondingly, the electron beam exhibits pronounced betatron oscillations in transverse momentum, the self-generated magnetic field is not yet able to sustain net pinching.

As the interaction proceeds, the tightly focused LG wake undergoes a nonlinear evolution in which the inner doughnut-bubble structure merges toward the axis, forming a dense electron population near the bubble center. As the LG pulse subsequently defocuses, these axis electrons expand outward and drift backward relative to the main bunch, generating a transient transverse kick to the off-axis high-energy electron bunches, visible as a pronounced electric-field contribution (red line) between $t\approx2220$ and $2464~\mathrm{fs}$ in Fig.~\ref{fig4}c. This kick redistributes the transverse phase space population and initiates the transition toward the magnetic pinching phase.
Following this transition, the self-generated magnetic field becomes the dominant factor governing the transverse dynamics. 
Figure~\ref{fig4}b illustrates the spatial correlation between the electron bunch and the magnetic field. The orange and purple contours denote electrons experiencing positive and negative magnetic forces, respectively. 
For both electron populations located above and below the bubble axis, the magnetic force is predominantly directed toward the axis. Consequently, the self-generated magnetic field acts as an effective pinching mechanism, continuously reducing the transverse displacement and momentum of the electrons. 
Through the $\mathbf{v}\times\mathbf{B}$ interaction, transverse momentum is progressively removed from the beam, leading to collective magnetic pinching and beam compression, accounting for the pronounced decrease of $u_y$ around $t=2700~\mathrm{fs}$ and the formation of a collimated beam, as observed in Fig.~\ref{fig3}d.

Figure~\ref{fig4}d shows the evolution of the particle-weighted magnetic pinching index,
$D_B=-\bigl\langle \theta_y F_{B,y}\bigr\rangle_w \big/ \bigl\langle |\theta_y F_{B,y}|\bigr\rangle_w,$
where $\langle\cdot\rangle_w$ denotes a particle-weighted average over the high-energy electrons.
Here, $D_B>0$ indicates net magnetic pinching. In the SMP case, $D_B$ rapidly increases and remains positive for an extended period, indicating that a large fraction of electrons enters a sustained pinching phase. In contrast, without SMP, the transition to positive $D_B$ is significantly delayed and persists for only a short duration, implying that the pinching phase is established much less effectively.

The origin of this difference can be understood from the evolution of the peak self-generated magnetic field shown in Fig.~\ref{fig4}e. In the SMP case, the wake evolution generates a transient transverse kick that redistributes the transverse phase space population and transfers a substantial fraction of electrons into the pinching region while the self-generated magnetic field remains strong. This process corresponds to the orange-shaded interval highlighted in Fig.~\ref{fig4}c--e, during which the electron bunch undergoes rapid phase space redistribution. As a result, a large fraction of electrons enters the pinching phase before the magnetic field weakens.

Following the kick stage, the interaction enters the green-shaded interval in Fig.~\ref{fig4}c--e, where the total transverse power becomes negative and the beam undergoes net transverse pinching. During this stage, the self-generated magnetic field continuously removes transverse momentum and drives collective beam compression. Without the synchronized kick, although the magnetic field remains sufficiently strong, only a limited fraction of electrons can access the pinching phase. Consequently, the beam occupies both pinching and anti-pinching regions of phase space, preventing sustained collective pinching and limiting the subsequent pinching process. These results demonstrate that efficient beam compression requires both a strong self-generated magnetic field and a pinching phase occupied by a substantial fraction of the beam. SMP fulfills both conditions by providing the phase space redistribution necessary for long-lived magnetic pinching.

\begin{figure}
    \centering
    \includegraphics[width=0.8\linewidth]{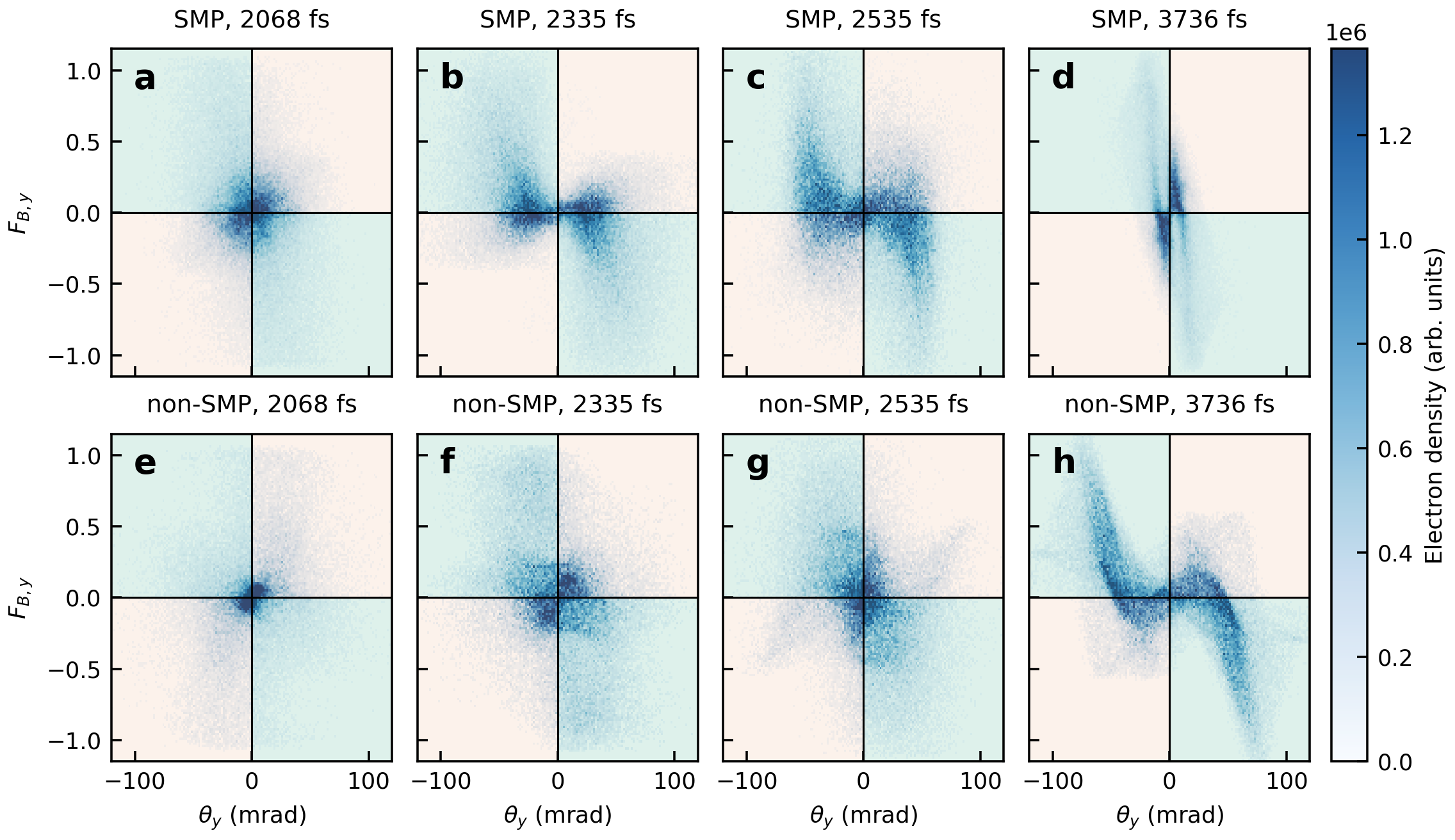}
    \caption{Kick-induced loading of electrons into the magnetic pinching phase. \textbf{a}--\textbf{d}, Electron density in $(\theta_y,F_{B,y})$ phase space for the cases with SMP at 2068, 2335, 2535 and 3736 fs. The peripheral-electron kick initiates a collective phase space redistribution that places most electrons in the pinching quadrants, where $F_{B,y}\theta_y<0$. \textbf{e}--\textbf{h}, Corresponding phase space evolution for non-SMP condition, where the kick-induced loading is absent or mistimed and the distribution does not develop dominant pinching-phase occupation.}
    \label{fig5}
\end{figure}

\textit{Transient kick enables SMP.}
Figure~\ref{fig5} further compares the phase space distributions in the $(\theta_y,F_{B,y})$ plane for the cases with and without SMP at four representative times. In this phase space, electrons satisfying $\theta_yF_{B,y}<0$ experience magnetic pinching and  $\theta_yF_{B,y}>0$ corresponds to magnetic anti-pinching. At $t=2068~\mathrm{fs}$, before the onset of the inner electron sheath kick, the electron distributions in both cases remain broadly distributed across the pinching and anti-pinching regions, indicating that no stable pinching phase has yet formed. At $t=2335~\mathrm{fs}$, shortly after the kick occurs in the SMP case, a substantial fraction of electrons is transferred into the pinching region. This redistribution is absent in the non-SMP case, where the electron population remains more evenly distributed between pinching and anti-pinching phases.

The distinction becomes even more pronounced at $t=2535~\mathrm{fs}$. In the SMP case, electrons accumulate predominantly in the pinching region, demonstrating the establishment of a stable magnetic pinching phase. Consequently, the self-generated magnetic field continuously extracts transverse momentum from a large fraction of the beam. In contrast, the non-SMP case does not exhibit comparable phase space trapping in the pinching region, even though a finite magnetic field is still present. These results confirm that the role of the inner-electron-induced kick is not merely to enhance the magnetic field interaction, but to transfer electrons into a stable pinching phase while the magnetic field remains strong. The combination of a strong self-generated magnetic field and a stable pinching phase enables sustained magnetic pinching and ultimately produces the compressed low-divergence beam.

\section*{Forming condition of the SMP}
The onset of SMP can be understood from the transverse equilibrium of electrons in the LG-driven wake. Electrons acquire orbital angular momentum from the twisted laser field, which introduces an effective centrifugal motion, while the ion-channel focusing field provides a restoring force toward the axis. The resulting equilibrium radius scales as \(r_{\mathrm{eq}}^2 \propto L_z/(\sqrt{\gamma}\,\omega_p)\), where \(L_z\) is the electron angular momentum and \(\omega_p\) is the plasma frequency. Using \(L_z\propto \ell a_0\), \(\gamma\propto n_e^{1/2}\), and \(\omega_p\propto n_e^{1/2}\), we obtain \(r_{\mathrm{eq}}^2\propto \ell a_0 n_e^{-3/4}\). We therefore define the normalized forming condition \(S\equiv (\ell a_0/6)(n_e/n_0)^{-3/4}\), where \(n_0=7.0\times10^{18}\,\mathrm{cm^{-3}}\) is the baseline plasma density adopted throughout this work. 
Using the present operating point \(\ell=1\) and \(a_0\simeq6\), the normalized forming condition parameter can be equivalently written as $
S \approx 0.717\,\ell a_0\,\left[n_e(10^{18}~\mathrm{cm^{-3}})\right]^{-3/4}$,
where \(n_e\) is expressed in units of \(10^{18}~\mathrm{cm^{-3}}\).
Simulations over a wide range of laser and plasma parameters show that SMP is realized when the interaction evolves with \(S\approx1\). The normalized forming condition therefore provides a practical criterion for preserving the transverse electron dynamics required for SMP under different laser and plasma conditions.

\textit{Transverse pinching contributes to energy gain.}
The same pinching process also contributes to longitudinal momentum evolution. In the relativistic limit, $u_z=\gamma v_z/c$ and $\gamma=\sqrt{1+u_\perp^2+u_z^2}$. Taking the time derivative gives $\mathrm{d}u_z/\mathrm{d}t=(\gamma/u_z)\,\mathrm{d}\gamma/\mathrm{d}t-(u_\perp/u_z)\,\mathrm{d}u_\perp/\mathrm{d}t$. The first term, $C=(\gamma/u_z)\,\mathrm{d}\gamma/\mathrm{d}t$, represents direct energy gain from the longitudinal electric field. The second term, $D=-(u_\perp/u_z)\,\mathrm{d}u_\perp/\mathrm{d}t$, represents the transverse pinching channel. Thus, the longitudinal momentum evolution can be approximated as $\mathrm{d}u_z/\mathrm{d}t\approx C+D$. Including the transverse pinching term improves the residual agreement by about 30\%, confirming that transverse momentum pinching contributes to the observed longitudinal energy gain. A detailed comparison is provided in Supplementary Fig.~S4. This interpretation is consistent with the enhanced electron energy in Fig.~\ref{fig2}d.

SMP operates in an intermediate-density regime between conventional LWFA and DLA. In this regime, strong plasma current enables the generation of a substantial self-generated magnetic field while avoiding the rapid dephasing characteristic of DLA. Unlike previously proposed autoresonant acceleration schemes that require externally applied or near-critical-density magnetic fields~\cite{liuGeneratingOvercriticalDense2013a}, SMP exploits the intrinsic nonlinear evolution of a wakefield driven by a twisted laser. The resulting $\mathbf{v}\times\mathbf{B}$ interaction transfers electrons into a stable magnetic pinching phase, continuously reducing transverse momentum and driving collective beam pinching. Consequently, SMP enables the production of high-charge electron beams with simultaneously reduced divergence.

At a fixed interaction plane located $z = 0.4 ~\mathrm{mm}$ downstream of the source, we define a geometric effective electron density $n_{\mathrm{eff}}=(Q/e)/[\pi(\theta_{\mathrm{rms}}z)^2\sigma_z]$, where $Q$ is the beam charge, $\theta_{\mathrm{rms}}=\theta_{\mathrm{FWHM}}/2.355$ is the root-mean-square (RMS) divergence angle assuming a Gaussian transverse profile, $z$ is the propagation distance from the source to the interaction plane, and $\sigma_z$ is the RMS bunch length. Using the experimentally measured parameters
($Q = 497$ pC, $\theta_{\mathrm{FWHM}} = 50$ mrad, $\sigma_z \approx
3~\mu\mathrm{m}$ from simulations), we obtain $n_{\mathrm{eff}} \approx
4.5 \times 10^{18}$ cm$^{-3}$ for the LG-driven beam and $n_{\mathrm{eff}}
\approx 6.4 \times 10^{17}$ cm$^{-3}$ for the Gaussian-driven beam---an
approximately sevenfold enhancement. This ratio is insensitive to the
choice of FWHM versus RMS definition, as the same conversion factor
applies to both cases. The substantial enhancement in effective density
highlights the potential of SMP for applications requiring high-flux
relativistic electron beams, including rare nuclear reaction studies and
laboratory astrophysics, while we note that for applications primarily
sensitive to total charge, the Gaussian driver may remain preferable.

\section*{Discussion}

The present results demonstrate that relativistic twisted light can actively regulate the transverse phase space of LPA through the SMP mechanism.  Distinct from the recently demonstrated LG-driven DLA mechanism~\cite{wangExperimentalDemonstrationDirectional2026}, which employs circularly polarized LG pulses interacting with overdense plasmas ($n_e \approx 3.2 \times 10^{22} \mathrm{cm}^{-3}$), where the incident laser is reflected from the plasma surface and beam collimation is sustained through direct coupling with the reflected laser field, the present SMP mechanism operates in underdense plasmas ($n_e \approx 7 \times 10^{18} \mathrm{cm}^{-3}$). Such an overdense plasma regime cannot satisfy the conditions required for SMP formation and its associated beam collimation.  A detailed comparison between the two mechanisms is provided in Supplementary. In the present regime, the LG mode primarily initiates the collective beam evolution rather than directly providing longitudinal acceleration. Specifically, the structured plasma current generates a quasi-static azimuthal magnetic field, while the nonlinear evolution of the wake produces a dense inner electron sheath. As the laser subsequently defocuses, this sheath expands outward and delivers a transient transverse kick that transfers off-axis high-energy electrons into a magnetic pinching phase. The resulting $\mathbf{v}\times\mathbf{B}$ interaction continuously damps the transverse momentum, transforming an initially separated two-lobe distribution into a dense, well-collimated electron beam. Unlike conventional LWFA, where increasing beam charge is generally accompanied by increased beam divergence, SMP simultaneously enhances beam collimation and effective electron-beam density during the acceleration process itself, without relying on external beam transport or post-acceleration collimation.

The forming condition further suggests that SMP can be extended to higher-power laser systems by jointly increasing the plasma density and the orbital angular momentum of the driver. Consistent with this forming condition, simulations with $\ell=3$ at $1~\mathrm{PW}$ and $n_e\simeq2.3\times10^{19}~\mathrm{cm}^{-3}$ exhibit electron charges of $\sim3.7~\mathrm{nC}$ above $10~\mathrm{MeV}$ while maintaining a divergence of only $\sim60~\mathrm{mrad}$, corresponding to an effective density of $n_e^{\mathrm{eff}}\approx3\times10^{19}~\mathrm{cm}^{-3}$. Simulations at $4~\mathrm{PW}$ further indicate that the density enhancement enabled by SMP can persist well beyond the present experimental conditions. These results suggest that SMP may provide a route toward ultra-dense relativistic electron beams in future multi-petawatt laser facilities. We caution that the present experimental data are obtained at a single point
in the $(n_e,\ell)$ parameter space; direct experimental validation of the
forming condition across varied plasma densities and OAM modes remains an
important direction for future work. Based on the forming condition, we predict that a
10 PW-class laser operating with $\ell = 3$ at $n_e \sim 2 \times 10^{19}
~\mathrm{cm}^{-3}$ should exhibit the characteristic SMP signatures---
divergence collapse and sustained magnetic pinching phase formation---on time
scales accessible to existing diagnostic techniques.

The results also clarify the physical conditions required for magnetic pinching. A strong self-generated magnetic field alone is insufficient, and neither is the presence of a transverse electron kick. Efficient pinching requires the simultaneous fulfillment of two conditions: a strong magnetic field and a stable magnetic pinching phase occupied by a substantial fraction of the beam. The comparison between the optimized and non-optimized cases shows that the kick must occur while the magnetic field remains near its maximum strength, thereby enabling sustained pinching over an extended interval. The mechanism therefore relies on the combined action of LG-driven current formation, inner electron sheath induced phase space redistribution, magnetic pinching, and collective beam pinching.

Although the present results establish SMP under experimentally accessible conditions, its ultimate performance at higher laser powers and larger OAM modes remains to be explored. The forming condition analysis suggests a pathway toward substantially higher beam charges and effective densities in future multi-petawatt laser systems. Furthermore, the efficiency of beam compression depends sensitively on the synchronization between kick-induced phase space loading and the persistence of the self-generated magnetic field, highlighting additional opportunities for optimization through laser-mode engineering and plasma tailoring.

More broadly, these results establish optical mode structure as an active control parameter for LPA. Rather than treating beam divergence as a by-product to be corrected after acceleration, the laser mode itself can be engineered to generate internal electromagnetic forces that reshape the beam during acceleration. The demonstrated enhancement of beam density and collimation is promising for applications requiring high-flux relativistic electron beams, including compact $\gamma$ rays sources~\cite{xuGenerationTerawattAttosecond2021}, high-flux neutron generation, rare nuclear interaction studies~\cite{guntherForwardlookingInsightsLasergenerated2022}, and laboratory astrophysics. Further optimization of laser mode structure, plasma composition, and guiding conditions may enable even higher-density beams and extend this approach toward next-generation high-brightness particle and radiation sources.

\section*{Acknowledgements}
This work was supported by the National Key Research and Development Program of China (Grants No. 2021YFA1601700, No. 2024YFA1610900, and No. 2025YFF0515101), the National Natural Science Foundation of China (Grants No. 12425510, No. U2267204, No. 12441506, No. 12575262, and No. 12388102), the Fundamental and Interdisciplinary Disciplines Breakthrough Plan of the Ministry of Education of China (Grant No. JYB2025XDXM204), the AI for Science Program, Shanghai Municipal Commission of Economy and Informatization (Grant No. 2025-GZL-RGZN-BTBX02029), the Strategic Priority Research Program of the Chinese Academy of Sciences (Grant No. XDA0380000), and the Natural Science Foundation of Shanghai (Grant No. 25JD1404100). The computations in this paper were performed on the $\pi$ 2.0 and Siyuan clusters supported by the Center for High Performance Computing at Shanghai Jiao Tong University. The authors also acknowledge support from the Yangyang Development Fund, China.

\bibliographystyle{naturemag-doi}
\bibliography{mycite}

\end{document}